\documentclass[aps,prl,reprint,superscriptaddress]{revtex4-2}
\usepackage{graphicx}
\usepackage{amsmath}
\usepackage{hyperref}

\begin{document}

\title{Self-organization drives symmetry-breaking, scaling, and critical growth transitions in stem cell-derived organoids}

\author{Daniel Aguilar-Hidalgo}
\affiliation{School of Biomedical Engineering, University of British Columbia, Vancouver, British Columbia, Canada}
\affiliation{Michael Smith Laboratories, University of British Columbia, Vancouver, British Columbia, Canada}

\author{Joel Ostblom}
\affiliation{Department of Statistics, University of British Columbia, Vancouver, BC, Canada}

\author{M Mona Siu}
\affiliation{School of Biomedical Engineering, University of British Columbia, Vancouver, British Columbia, Canada}
\affiliation{Department of Medical Genetics, University of British Columbia, Vancouver, British Columbia, Canada}

\author{Divy Raval}
\author{Ajinkya Ghagre}
\author{Tiam Heydari}
\author{Benjamin McMaster}
\author{Jonathan Gui}
\affiliation{School of Biomedical Engineering, University of British Columbia, Vancouver, British Columbia, Canada}
\affiliation{Michael Smith Laboratories, University of British Columbia, Vancouver, British Columbia, Canada}

\author{Nicolas Werschler}
\affiliation{School of Biomedical Engineering, University of British Columbia, Vancouver, British Columbia, Canada}

\author{Mukul Tewary}
\affiliation{NHS Blood and Transplant, London, England}

\author{Peter W. Zandstra}
\affiliation{School of Biomedical Engineering, University of British Columbia, Vancouver, British Columbia, Canada}
\affiliation{Michael Smith Laboratories, University of British Columbia, Vancouver, British Columbia, Canada}

\begin{abstract}
The emergence of spatial patterns and organized growth is a hallmark of developing tissues. While symmetry-breaking and scaling laws govern these processes, how cells coordinate spatial patterning with size regulation remains unclear. Here, we combine quantitative imaging, a Turing activator–repressor model with self-organized reactive boundaries, and in vitro models of early mouse development to study mesodermal pattern formation in two-dimensional (2D) gastruloids. We show that colony size dictates symmetry: small colonies (radius $\sim$100\,µm) spontaneously break symmetry, while larger ones remain centro-symmetric, consistent with size-dependent positional information and model predictions. The mesodermal domain area scales robustly with colony size following a power law, independent of cell density, indicating that cells sense and respond to gastruloid size. Time-lapse imaging reveals a biphasic growth law: an early power-law expansion followed by exponential arrest, marking a dynamical phase transition. These dynamics, conserved across sizes, reflect features of criticality seen in physical systems, where self-organization, scaling, and boundary feedback converge. Our findings uncover a minimal mechanism for size-dependent pattern formation and growth control. This framework enables quantitative investigation of symmetry-breaking and scaling in self-organizing tissues, offering insights into the physical principles underlying multicellular organization.
\end{abstract}

\maketitle

\section*{Introduction}

From a single fertilized cell, complex organisms emerge through a series of coordinated events involving pattern formation, symmetry-breaking, and regulated growth. Developmental systems exhibit striking parallels with physical systems, where global organization arises from local interactions and feedback. In particular, they display behaviors reminiscent of critical transitions—where small perturbations can induce large-scale reorganizations—as observed near phase transitions in condensed matter and statistical physics~\cite{bak_complexity_1995,zillio_incipient_2008,valverde_structural_2015,depersin_global_2018,munoz_colloquium_2018,lenne_sculpting_2022}. Understanding whether, and how developmental patterning operates near such critical points could yield fundamental insights into the physics of biological self-organization.

Organoids are self-organized cellular aggregates that recapitulate aspects of \textit{in vivo} development and have emerged as powerful model systems for probing otherwise inaccessible stages of early embryogenesis and identifying minimal components of tissue-level organization~\cite{lewis_signals_2008}. Their tractability, reproducibility, and amenability to perturbation have enabled investigations of phenomena such as cortical layer formation~\cite{lancaster_cerebral_2013}, crypt-villus architecture~\cite{serra_self-organization_2019}, and axial patterning~\cite{lawlor_cellular_2021}. Yet, despite these advances, it remains unclear how organoid systems encode and regulate the size, shape, and symmetry of emergent cellular patterns. This is especially true under constraints such as geometry, dimensionality, or external signaling.

A central open question is whether symmetry-breaking and size-scaling emerge intrinsically or require higher-dimensional tissue organization. While three-dimensional (3D) gastruloids have demonstrated spontaneous axial polarization reminiscent of anterior–posterior patterning~\cite{turner_anteroposterior_2017,van_den_brink_single-cell_2020,merle_precise_2023,anlas_early_2024}, such asymmetries have not been reported in two-dimensional (2D) adherent gastruloids. This raises the fundamental question: is dimensionality essential for spatial asymmetry to emerge, or can intrinsic self-organizing mechanisms also operate in 2D?

Here, we integrate high-content imaging, quantitative analysis, and theoretical modeling to investigate the emergence of mesodermal patterns in 2D gastruloids—pluripotent stem cell colonies cultured on micropatterned substrates under differentiation-inducing conditions~\cite{harrison_assembly_2017,morgani_micropattern_2018,amadei_embryo_2022}. Our data from over \(10^4\) colonies reveal that symmetry-breaking of mesodermal domains occurs robustly in small-radius colonies, challenging the view that 3D organization is a prerequisite for axis formation. Furthermore, we uncover a robust power-law scaling between pattern size and colony size, and show that this relation persists under density perturbations, suggesting that cells possess an intrinsic mechanism for sensing size and geometry.

Live imaging of domain expansion dynamics reveals a biphasic growth law: an initial power-law regime transitions into exponential relaxation, consistent with a critical slowing down followed by growth arrest. This growth control regime emerges at a shared transition window across colony sizes, consistent with scale-invariant behavior near a critical threshold. These findings are captured by a Turing-type activator–repressor model with self-organized reactive boundaries, which couples morphogen patterning to boundary interactions without requiring external symmetry-breaking inputs. 

Together, our results reveal that 2D gastruloids exhibit spontaneous symmetry-breaking, robust scaling, and critical transition dynamics—hallmarks of self-organized developmental systems operating near criticality. This work provides a minimal framework to explore the emergence of biological structure from physical principles, and opens new directions for understanding how size, geometry, and boundary effects shape multicellular organization.


\section{Size-dependent symmetry-breaking in 2D gastruloids}

To investigate the dynamics, scaling, and growth behaviors of cellular patterns in organoids, we used 2D adherent pluripotent stem cell (PSC)-derived gastruloids, which offer high reproducibility and robustness while enabling high-throughput quantification. As a readout, we measured the size of the mesodermal cellular domain, characterized by Brachyury (Bra) expression, and its location within the gastruloid. Specifically, we focused on the emergence of polarized or asymmetric patterns reminiscent of \textit{in vivo} symmetry-breaking events that lead to the formation of the anterior-posterior (AP) axis. Our control parameters included gastruloid area and initial seeding cell density.

\begin{figure*}[]
    \centering
    \includegraphics[width=0.9\textwidth]{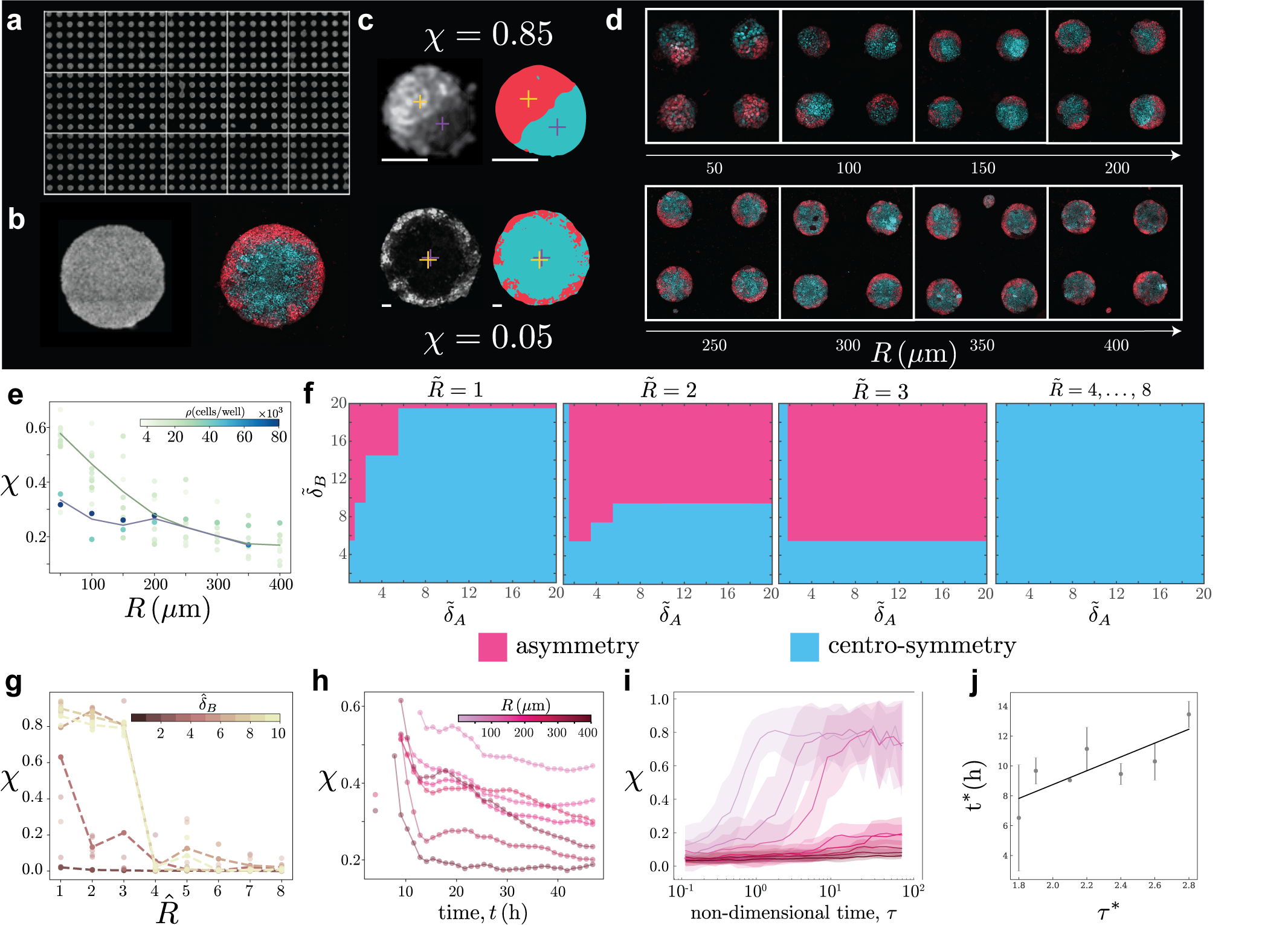} 
    \caption{\textbf{Size-dependent symmetry-breaking in 2D PSC colonies is captured by a Turing model with self-organized boundaries.} 
(a) Bright-field image of high-throughput PSC colony differentiation on a micro-patterned substrate. 
(b) Confocal image at 48\,h showing marker expression: Brachyury (Bra, red), Sox2 (cyan); $R=400\,\mu$m. 
(c) Segmentation of Bra$^+$ (red) and Bra$^-$ (cyan) regions with centroids marked (yellow/purple crosses). Scale bar: $50\,\mu$m. 
(d) Representative Bra$^+$ patterns after 48\,h: smaller colonies ($R=50{-}150\,\mu$m) exhibit symmetry-breaking, while larger ones ($R=200{-}400\,\mu$m) remain centro-symmetric. 
(e) Symmetry score $\chi$ for colonies of $R = 50{-}400\,\mu$m and seeding densities $\rho = 4{-}80 \cdot 10^3$ cells/well. Small colonies show $\rho$-dependent symmetry-breaking. N=15305.
(f) Phase diagram of simulated colonies showing symmetry regimes as a function of normalized radius $\hat{R}=R/\lambda$ and diffusive penetration length $\hat{\delta}_i = D_i / (\sigma_i \lambda)$. Pink: symmetry-breaking; blue: centro-symmetry. 
(g) Simulated $\chi$ as a function of $\hat{R}$ shows dependence on $\hat{\delta}_B$. 
(h) Experimental $\chi(t)$ shows rapid early relaxation across colony sizes. N=520.
(i) Simulated symmetry score $\chi(\tau)$ for $\hat{R}=1{-}8$ shows time evolution toward steady-state; shaded bands: standard deviation across replicates. 
(j) Experimental symmetry decision time $t^*$ vs. simulated time $\tau^*$ shows a consistent scaling relation under $\lambda = 50\,\mu$m. Linear fit: $\tau^* = 0.14t^* - 0.19$ ($R^2 = 0.64$); $k_A = 0.14 \pm 0.12$ h$^{-1}$. Error bars: standard deviation.
}
    \label{fig:symmetry}
\end{figure*}

To control the gastruloid area, we used micropatterning technology~\cite{lee_micropatterning_2009,warmflash_method_2014}, which allowed us to pre-print regions with specific sizes and shapes onto which PSCs were seeded. The PSC colony, confined to the micropatterned region, approximately maintained its size and shape during the studied developmental period (48 hours post-induction) (\textbf{Fig.~\ref{fig:symmetry}a,b}). We explored circular geometries with radii ranging from $50$ to $400~\mu\mathrm{m}$, encompassing the anterior-posterior (A--P) axis length of the actual embryo during development~\cite{morgani_micropattern_2018}, and seeding cell densities ($\rho$) ranging from $4 \times 10^3$ to $80 \times 10^3~\mathrm{cells/well}$. Upon differentiation media induction (see Methods), colonies formed a defined mesodermal domain within 48 hours (\textbf{Fig.~1b}).

To determine whether 2D adherent gastruloids exhibit mesodermal domain symmetry-breaking, we quantified the surface area of the Bra$^+$ pattern, $S_p$, and its location within the entire cellular colony, $S_c$. We defined a symmetry score, $\chi$, as the normalized distance between the Bra-weighted centroid of the mesodermal domain and the centroid of the Bra$^-$ region (\textbf{Fig.~\ref{fig:symmetry}c}; see Supplementary Information for details). A perfectly symmetric domain centered within the gastruloid yields $\chi = 0$, while a small off-center domain at the boundary corresponds to $\chi = 1$. A notable intermediate value, $\chi = 8/(3\pi)$, reflects a domain occupying exactly half the colony area, indicating bilateral symmetry.

We found that small gastruloids ($R = 50{-}150~\mu\mathrm{m}$) exhibited high symmetry scores ($\chi = 0.6 \pm 0.2$), consistent with bilateral symmetry. Larger gastruloids ($R > 200~\mu\mathrm{m}$) exhibited lower scores ($\chi = 0.2 \pm 0.2$), reflecting centro-symmetric patterns (\textbf{Fig.~\ref{fig:symmetry}d,e}). This size-dependent transition emerged most clearly at low seeding densities ($\rho < 40 \times 10^3~\mathrm{cells/well}$). At higher densities, even small gastruloids shifted toward centro-symmetric configurations, suggesting that cell density may modulate the system's effective diffusivity or boundary interactions. Together, these results point to an underlying mechanism for size- and density-dependent symmetry-breaking in mesodermal patterning.

Centro-symmetric patterns in adherent gastruloids have previously been explained by Turing-like activator-repressor dynamics~\cite{deglincerti_self-organization_2016,etoc_balance_2016,tewary_stepwise_2017,tewary_high-throughput_2019,kaul_virtual_2023}, while bilateral patterns have been associated with asymmetrically imposed boundary fluxes~\cite{manfrin_engineered_2019}. These models typically assume static or externally imposed boundary conditions. In contrast, we explore the effects of \textit{reactive boundaries}, where morphogen flux depends on local concentrations~\cite{brauns_bulk-surface_2021,burkart_dimensionality_2024}. Such boundaries introduce dynamic feedback between the bulk and edge, allowing positional information to emerge intrinsically and enabling pattern adaptation with system size.

We hypothesized that a minimal two-component Turing model with reactive boundary conditions can recapitulate these effects and explain the observed scaling behavior. In this framework, the system's size and seeding density modulate the effective diffusion coefficients of morphogens, altering pattern length scales and diffusive penetration lengths.

To test this, we implemented a two-component reaction-diffusion model with a sharp-switch production term and concentration-dependent boundary fluxes~\cite{turing_alan_mathison_chemical_1952,dillon_pattern_1994,erban_reactive_2007,werner_scaling_2015,wurthner_bridging_2022}. The bulk dynamics are governed by:
\begin{align}
    \partial_t A &= \nabla D_A \nabla A - k_A A + \nu_A P, \label{eq:activator} \\
    \partial_t B &= \nabla D_B \nabla B - k_B B + \nu_B P, \label{eq:repressor}
\end{align}
where $A$ and $B$ are the concentrations of activator and repressor; $D_i$, $k_i$, and $\nu_i$ denote diffusion, degradation, and production rates, respectively. The production function $P(A, B) = A^h / (A^h + B^h)$ acts as a binary switch for large $h$, setting $P=1$ when $A>B$, and $P=0$ otherwise~\cite{werner_scaling_2015}. Boundary fluxes follow:
\begin{align}
    D_A \nabla A \cdot \hat{n} &= \sigma_A A, \label{eq:boundary_a} \\
    D_B \nabla B \cdot \hat{n} &= \sigma_B B. \label{eq:boundary_b}
\end{align}

We simulated pattern formation using non-dimensional parameters: $\hat{R} = R/\lambda$ (with $\lambda = \sqrt{D_A / k_A}$), and $\hat{\delta}_i = D_i / (\sigma_i \lambda)$ representing normalized diffusive penetration lengths. From here onward, hat notation denotes non-dimensional quantities. Simulations were initialized with low-amplitude random fluctuations of $A$ and $B$ and run to steady state.

The model robustly reproduced key experimental trends. It predicted symmetry-breaking patterns for $\hat{R} = 1{-}3$, and centro-symmetric configurations for larger $\hat{R}$ or smaller $\hat{\delta}_i$ values (\textbf{Fig.~\ref{fig:symmetry}f--g}). These results match experimental data assuming $\lambda = 50~\mu\mathrm{m}$, consistent with reported morphogen diffusion length scales. In boundary-dominated regimes (low $\hat{\delta}_i$), the system remained centro-symmetric even for small $\hat{R}$, aligning with high-density experimental conditions.

To evaluate whether the model captures not just steady-state patterns but also dynamic features, we next analyzed the temporal evolution of the symmetry score $\chi$ in both simulations and live-imaging data. Experimental colonies stabilized their symmetry configuration within $t \approx 12{-}16$ hours post-induction (\textbf{Fig.~1h}). Simulations showed analogous trends in $\chi(\tau)$ across $\hat{R}$ values (\textbf{Fig.~\ref{fig:symmetry}i}), enabling estimation of a conversion factor $k_A = (0.14 \pm 0.12)\,\mathrm{h}^{-1}$ between experimental and model timescales (\textbf{Fig.~\ref{fig:symmetry}j}).

Interestingly, the dynamics of the simulated mesodermal pattern $\hat{S}_p$ closely matched those of the production domain $\hat{S}_s$, defined by:
\begin{equation}
    \hat{S}_s(\tau) = \int_{\Omega} P(\hat{x}, \hat{y}, \tau)\, d\hat{S},
\end{equation}
where \( d\hat{S} \) is the infinitesimal area element over the 2D non-dimensional domain \( \Omega \). This overlap reveals a key insight: that the morphogen production region itself — not just the resulting concentration field — may drive fate specification and spatial organization. These results suggest that morphogen sources act as dynamic patterning agents, potentially guiding symmetry-breaking through their emergence, shape, and temporal evolution — a concept also observed in the self-organized expansion of signaling domains during early \textit{Drosophila} development~\cite{aguilar-hidalgo_critical_2018}. 

Given that both the symmetry of the pattern and the identity of its morphogen source depend on system size, we next asked whether the \textit{extent} of these domains also scales with colony size, and whether this scaling follows a universal growth law. If so, this would suggest that the same self-organizing principles that regulate pattern symmetry dynamically govern the spatial expansion of fate domains—linking symmetry-breaking, scaling, and growth within a unified framework.

\section{Emergent Scaling Laws in Mesodermal Pattern Formation}

Indeed, both simulations and experiments revealed a striking scaling relationship between the mesodermal domain area (\( S_p,\, \hat{S}_p \)) and the overall colony area (\( S_c,\, \hat{S}_c \)), suggesting that pattern geometry grows proportionally with system size. Smaller colonies produced visibly constrained domains, while larger colonies exhibited expanded fate territories (\textbf{Fig.~2a}). These observations prompted a quantitative investigation of the scaling behavior, aiming to determine whether a universal law governs the growth of fate domains across conditions.

From our simulation results, we analyzed the relationship between the normalized mesodermal pattern surface area \( \hat{S}_p \) and the normalized colony surface area \( \hat{S}_c \) across a range of diffusive penetration lengths \( \hat{\delta}_i \). We found that these quantities follow a power-law scaling relation that defines an emergent \textit{mesodermal pattern scaling law}:
\begin{equation}
    \hat{S}_p \sim \hat{S}_c^\alpha,
\end{equation}
where \( \alpha \) is the scaling exponent (\textbf{Fig.~2}). We also investigated the scaling of the morphogen source area \( \hat{S}_s \) with respect to colony size. Both \( \hat{S}_p \) and \( \hat{S}_s \) exhibit power-law scaling, although the latter is slightly noisier and characterized by lower scaling exponents (\textbf{Fig.~2b,c}).

By systematically scanning values of \( \hat{\delta}_A \) and \( \hat{\delta}_B \), we observed that the scaling exponent \( \alpha \) is largely insensitive to changes in \( \hat{\delta}_A \) (\textbf{Sup. Fig. 1}), but strongly dependent on \( \hat{\delta}_B \) (\textbf{Fig.~2d,e}). In particular, lower values of \( \hat{\delta}_B \) led to smaller initial values of \( \alpha \) that progressively increased over time, reaching quasi-linear values at steady state. In contrast, higher values of \( \hat{\delta}_B \) yielded moderate and time-invariant values of \( \alpha \). A similar trend was observed for the morphogen source area \( \hat{S}_s \), though with slightly lower steady-state values. These results suggest that the effective penetration depth of the repressor \( B \) controls the temporal evolution and final value of the scaling exponent, shaping how spatial domains scale with system size, whereas the activator’s diffusion length plays a negligible role in this scaling behavior.

\begin{figure*}[]
    \centering
    \includegraphics[width=0.9\textwidth]{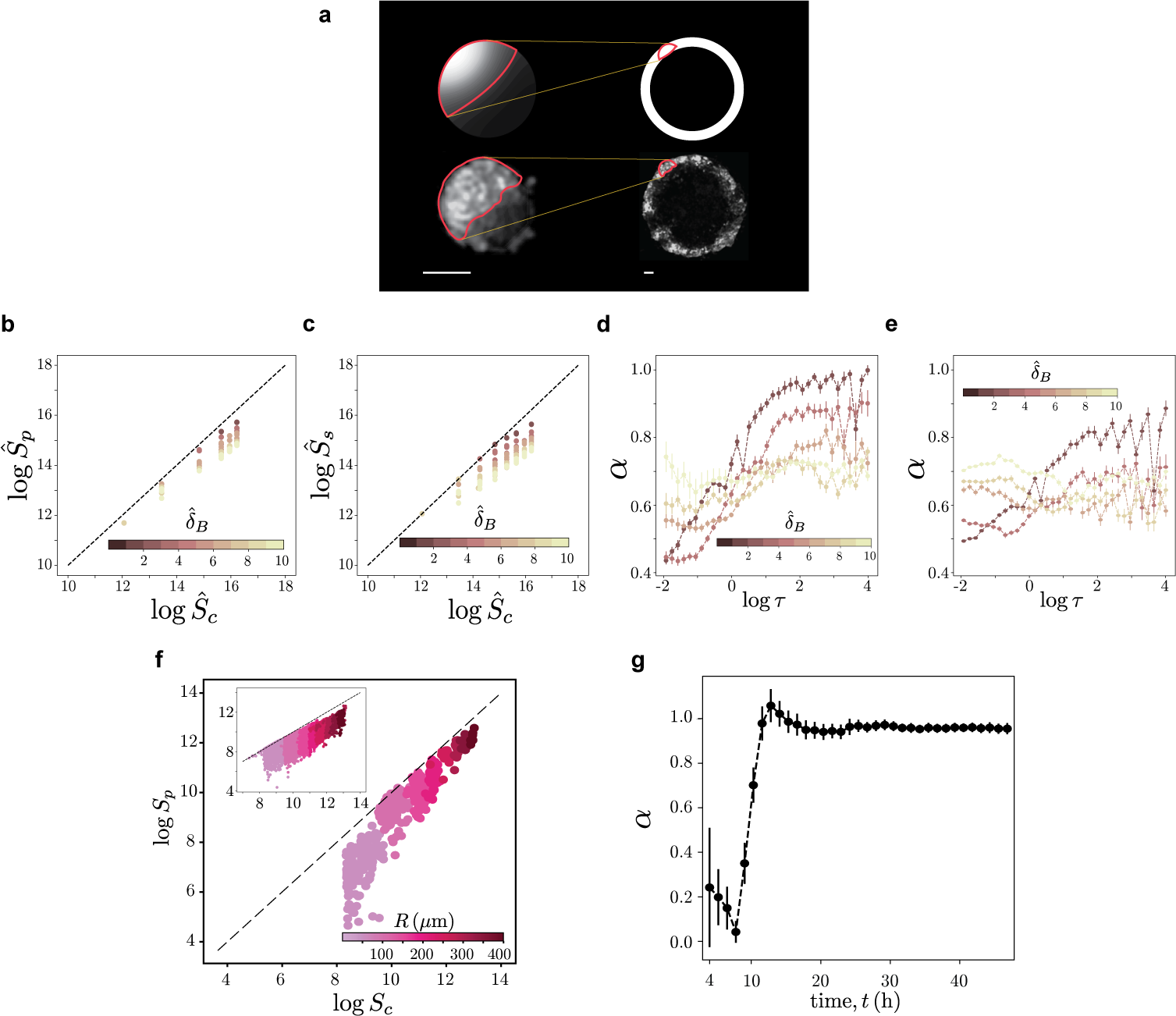} 
    \caption{\textbf{Scaling behavior of the mesodermal pattern agrees with model predictions.} 
\textbf{(a)} Simulated (top) and experimental (bottom) colonies showing size-dependent patterning of the mesodermal domain. Scale bar: $50~\mu$m. 
\textbf{(b)} Simulation results showing power-law scaling between normalized pattern area $\hat{S}_p$ and colony area $\hat{S}_c$, with scaling exponent $\alpha$ estimated separately for each condition. Color map denotes the variation of $\alpha$ as a function of the effective diffusive penetration length of the repressor, $\hat{\delta}_B$. 
\textbf{(c)} Analogous scaling between normalized morphogen source area $\hat{S}_s$ and colony area $\hat{S}_c$, with $\alpha$ also estimated per condition and shown as a function of $\hat{\delta}_B$. 
\textbf{(d)} Temporal evolution of the scaling exponent $\alpha$ for the mesodermal pattern area $\hat{S}_p$. Low values of $\hat{\delta}_B$ (e.g., 2, 4) exhibit a time-dependent increase in $\alpha$, converging to near-linear scaling ($\alpha \approx 0.98$ and $0.9$, respectively), while higher $\hat{\delta}_B$ values (e.g., 10) remain stable ($\alpha \approx 0.6\text{--}0.7$). A characteristic relaxation time of $\log \tau^* \approx 2$ marks the transition for low $\hat{\delta}_B$. 
\textbf{(e)} Dynamics of the scaling exponent $\alpha$ for the morphogen source area $\hat{S}_s$, showing similar but more variable behavior compared to $\hat{S}_p$, with slightly lower steady-state values (e.g., $\alpha \approx 0.8$ for $\hat{\delta}_B = 2$, and $\alpha \approx 0.7$ for $\hat{\delta}_B = 4$). 
\textbf{(f)} Experimental results confirm power-law scaling between pattern area $S_p$ and colony area $S_c$, with a measured exponent $\alpha = 0.95 \pm 0.03$ at steady state (48 h), based on live imaging of $N = 520$ colonies. Inset: Fixed colonies at 48 h ($N = 15{,}305$), spanning all seeding densities shown in Fig.~1 ($\rho = 4\text{--}80 \times 10^3$ cells/well), also exhibit consistent sublinear scaling, reinforcing the robustness of the quasilinear growth law across experimental conditions. 
\textbf{(g)} Temporal evolution of the experimental scaling exponent $\alpha$, quantified independently at each time point across $N=520$ colonies. Black dots indicate mean values, with error bars representing the standard deviation of the mean. In panels (d--e), simulated $\alpha$ values are averaged over different $\hat{\delta}_A$ values, with shaded bands showing standard error across $n=5$ independent runs per condition.}
\label{fig:scaling}

\end{figure*}

To compare these findings with our experimental results, we plotted the surface area of the Bra$^+$ domain (\( S_p \)) against the surface area of the respective gastruloid (\( S_c \)) (\textbf{Fig.~2f}). Remarkably, we found that these quantities followed the same type of scaling law as in Eq.~(6). This scaling law applied consistently across all experimental conditions, including variations in micropattern radius \( R \) and seeding cell density \( \rho \). The scaling exponent was found to be \( \alpha = 0.93 \pm 0.13 \), indicating a sublinear but nearly-linear relationship between \( S_p \) and \( S_c \) (\textbf{Fig.~2g}).

We observed that very small colonies (\( R \sim 50\, \mu\mathrm{m} \)) lag behind in the developmental process, misaligning with the overall scaling trend seen in the rest of the analyzed sizes. This deviation is likely due to insufficient cell numbers or ineffective morphogen diffusion in small colonies. The robust scaling behavior spanned two orders of magnitude in gastruloid size and a 20-fold range in seeding cell density, with a sample size of over \( 1.5 \times 10^4 \) colonies (\textbf{Fig.~2f} and \textbf{2f inset}).

These results suggest an adaptive mechanism that provides consistent positional information to cells across different domain sizes. When we compared our experimental data with the simulation results, we found that the experimentally determined scaling factors closely matched our model predictions (\textbf{Fig.~2b,c,f}). The best agreement occurred for intermediate values of \( \hat{\delta}_B \in [2,4] \), where simulations showed steady-state exponents of \( \alpha \approx 0.9 \), consistent with experiments. This suggests that the effective diffusive penetration depth of the repressor in experiments is approximately two- to four-fold the activator's decay length \( \lambda \).

Finally, to validate the biological plausibility of our model, we leveraged the time-dependent scaling exponent \( \alpha(t) \), which increased and stabilized by \( t^* \approx 12{-}16\, \mathrm{h} \) (\( \log{\tau^*} \approx 0.5{-}1 \); \textbf{Fig.~2d,e,g}). Matching this timescale between simulations and experiments enabled estimation of the degradation rate \( k_A = 0.14 \, \mathrm{h}^{-1} \) (\textbf{Fig.~1j}). Combined with the decay length \( \lambda = 50\,\mu\mathrm{m} \), previously identified in symmetry-breaking analysis (\textbf{Fig.~1}), this yields a diffusion coefficient \( D_A = 0.10 \, \mu\mathrm{m}^2/\mathrm{s} \)—within reported ranges for endogenous morphogens~\cite{kicheva_kinetics_2007, muller_differential_2012, zhou_free_2012, romanova-michaelides_morphogen_2022}. These results reinforce that dynamic scaling and growth control can emerge from physically grounded, self-organizing principles.

Beyond validating the model, this dynamic scaling reveals a transient overshoot in \( \alpha(t) \) in experimental data, not observed in simulations. This early deviation suggests a nonequilibrium adjustment period, possibly reflecting delayed coupling between morphogen production and diffusion. Such overshoot may transiently amplify positional information before settling into a stable scaling regime—a process reminiscent of refinement phases observed during early \textit{in vivo} patterning~\cite{gregor_stability_2007, wartlick_dynamics_2011, aguilar-hidalgo_hh-driven_2013-1, green_positional_2015}. That both scaling and symmetry stabilize around the same time highlights a shared underlying dynamic. This convergence invites the question: could the relaxation of scaling behavior signal a deeper dynamical transition? We explore this next.

\section{Power-law Growth and Critical Arrest in Mesodermal Pattern Dynamics}

The convergence of scaling and symmetry-breaking dynamics around a common relaxation time raises the possibility of a deeper organizing principle underlying growth. We therefore asked whether the approach to steady-state scaling reflects not just geometric adaptation, but a dynamical transition in growth behavior. The robust scaling of mesodermal domain size with colony radius suggests that cells access dynamic information about gastruloid size and their position relative to the boundary. In our model, this positional information arises from a Turing-like self-organizing mechanism with reactive boundaries, in which morphogen concentrations evolve toward steady-state profiles. As the activator \( A \) drives pattern formation and domain expansion, we hypothesize that the mesodermal domain area \( S_p \) follows a growth law that slows and ultimately arrests as morphogen levels saturate.

\begin{figure*}[]
    \centering
    \includegraphics[width=\textwidth]{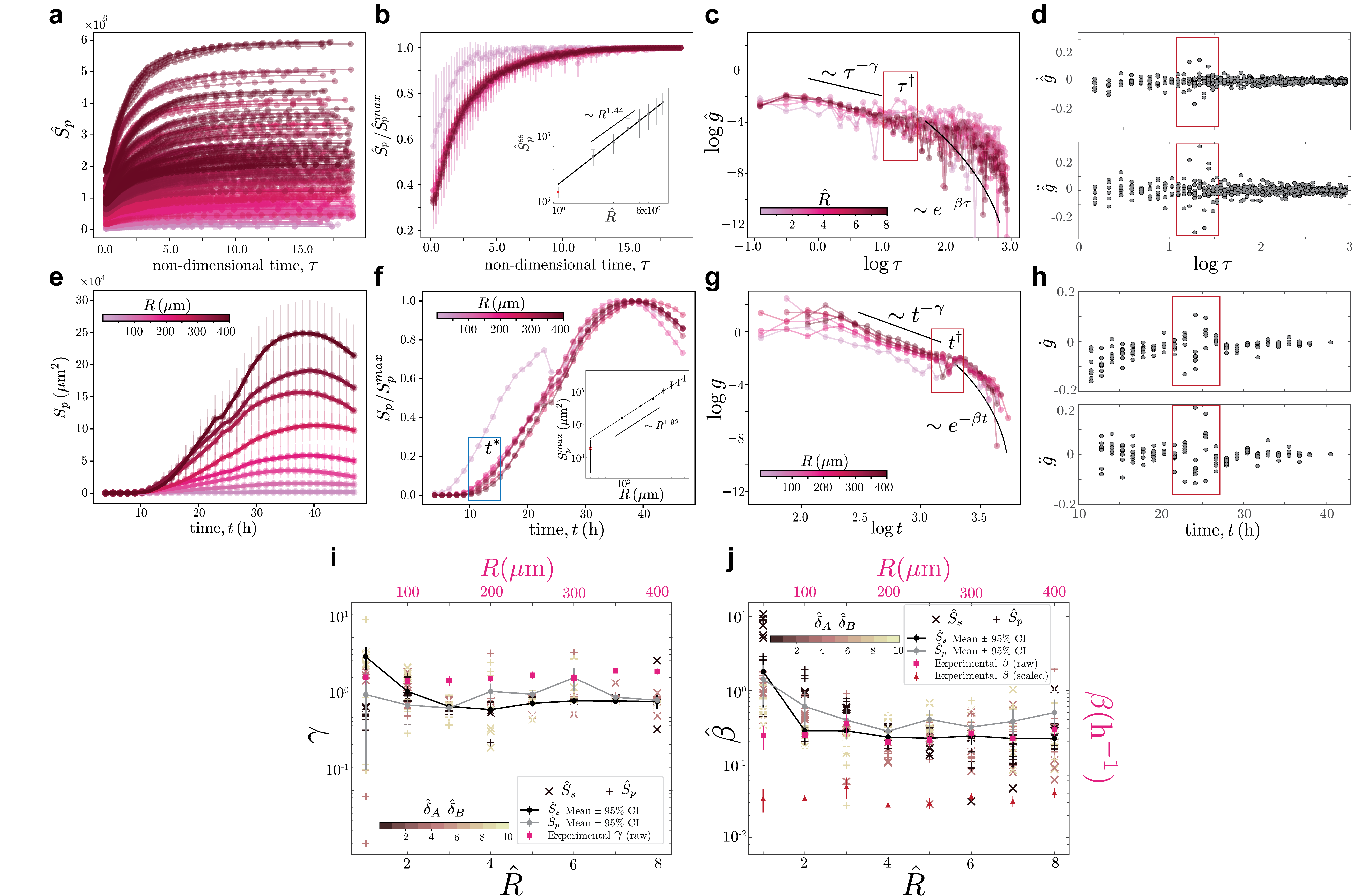} 
    \caption{\textbf{Pattern growth dynamics exhibit power-law decay followed by exponential tail, consistent with critical-like transitions predicted by the model.} 
(a) Dynamics of simulated pattern size for colony radii in the range $\hat{R}=1{-}8$. For each $\hat{R}$, curves represent averages across simulations with varying effective diffusion lengths $\hat{\delta}_A, \hat{\delta}_B \in [2, 10]$, sampled in steps of 2. 
(b) Normalized simulated pattern size trajectories collapse onto a master curve. 
Error bars indicate the standard error of the mean across all simulations per $\hat{R}$. 
Inset: Maximum pattern size reached at relaxation ($\dot{\hat{g}} = 0$) scales with colony radius via a power law, with exponent $\epsilon = 1.44 \pm 0.06$. 
(c) Simulated relative growth rate $\hat{g}(\tau)$ shows early-time power-law decay followed by an exponential tail across all system sizes. The transition occurs at $\log{\tau^{\dagger}} \sim 1.35\pm0.15$ (red box). 
(d) First and second time derivatives of $\hat{g}$ increase in amplitude near $\tau^{\dagger}$ (red box), consistent with critical-like behavior. 
(e) Experimental pattern size dynamics for colony radii $R = 50{-}400 \, \mu$m. Lines and markers show the mean per $R$, with error bars representing standard deviation across colonies. 
(f) Normalized experimental pattern sizes collapse onto a master curve. The blue box denotes the onset time $t^*$, marking the transition to power-law expansion across all conditions on (g).
Inset: Maximum pattern size at relaxation ($\dot{g} = 0$) scales with colony radius with exponent $\epsilon = 1.92 \pm 0.19$ (Mean $\pm$ 95\% CI). 
(g) Experimental growth rates $g(t)$ exhibit an early power-law decay ($\gamma$) followed by exponential relaxation ($\beta$), with a transition at $t^{\dagger} = 22{-}27$ h (red box). 
(h) First and second time derivatives of $g(t)$ increase in amplitude near $t^{\dagger}$ (red box), consistent with critical slowing down and a dynamical phase transition. 
(i,j) Growth exponents extracted from simulations and experiments: (i) power-law exponent $\gamma$; (j) exponential decay exponent $\beta$ (dimensional and non-dimensional). 
Exponents are estimated from $g(t)$ or $\hat{g}(\tau)$ trajectories. 
Symbols show individual replicate fits (source region $\hat{S}_s$: ×; non-source region $\hat{S}_p$: +). 
Means $\pm 95\%$ confidence intervals are shown in black and gray. Red squares: experimental exponent estimates aligned by $\hat{R}$ (assuming $\lambda = 50 \, \mu$m), with $\beta$ scaled using $k_A = 0.14 \, \text{h}^{-1}$. 
Panels (i, j) use dual x-axes: bottom x-axes show $\hat{R}$, top x-axes show physical $R$. 
Panel (j) additionally uses dual y-axes: left y-axis shows non-dimensional $\hat{\beta}$, right y-axis shows dimensional $\beta$ in units of h$^{-1}$. 
Experimental colonies: N = 520.
}
    \label{fig:critical}
\end{figure*}

To test this hypothesis, we analyzed the dynamics of pattern area \( \hat{S}_p(\tau) \) in simulations across colony sizes \( \hat{R} = 1\text{--}8 \). We observed that pattern growth follows a classic saturating trajectory in all cases, with the final size increasing with system size (\textbf{Fig.~\ref{fig:critical}a}). When normalized by their maximum values (defined by \(\dot{\hat{S}}_p = 0\)), the growth trajectories collapsed onto a master curve (\textbf{Fig.~\ref{fig:critical}b}). This collapse was robust to variations in the diffusive penetration lengths \( \hat{\delta}_A \) and \( \hat{\delta}_B \), which were sampled across \( [2, 10] \). The observed scaling collapse indicates that growth dynamics are self-similar across sizes, with minimal dependence on pattern symmetry or morphogen parameters.

Interestingly, the final pattern area in simulations scaled with colony radius as a power law: \( \hat{S}_p^{\max} \sim \hat{R}^{\epsilon} \) with exponent \( \epsilon = 1.44 \pm 0.06 \), consistent across parameter variations (\textbf{Fig.~\ref{fig:critical}b}, inset). This suggests that pattern growth is not only size-dependent but follows a scale-invariant law up to arrest. 

To further investigate growth dynamics, we computed the relative growth rate \( \hat{g} = \dot{\hat{S}}_p / \hat{S}_p \). Across all colony sizes, \( \hat{g}(\tau) \) exhibited two distinct dynamical regimes: an early-time power-law decay \( \hat{g} \sim \tau^{-\gamma} \), followed by an exponential tail \( \hat{g} \sim \exp(-\hat{\beta} \tau) \) (\textbf{Fig.~\ref{fig:critical}c}). The transition between these regimes occurred at a critical timescale \( \log{\tau^{\dagger}} \sim 1.35\pm0.15 \), approximately invariant with system size. 
Importantly, we found that the early-time growth exponent \( \gamma \)—which quantifies the rate of expansion during the initial power-law phase—exhibited a weak positive correlation with system size (\( r_s = 0.24 \), \( p = 9.3 \times 10^{-4} \)), suggesting that larger domains facilitate faster early pattern growth. In contrast, \( \gamma \) decreased more strongly with the effective repressor penetration length \( \hat{\delta}_B \) (\( r_s = -0.44 \), \( p = 5.3 \times 10^{-7} \)), indicating that greater inhibitory range slows early morphogenetic expansion (\textbf{Fig.~\ref{fig:critical}i}). These trends demonstrate how both the physical dimensions of the system and molecular-scale parameters modulate early pattern dynamics in the model. Conversely, the exponential decay exponent \( \hat{\beta} \), which characterizes the rate of growth arrest at later times, was largely independent of both \( \hat{R} \) and \( \hat{\delta}_B \), pointing to a relaxation process that is robust across system scales and parameter regimes (\textbf{Fig.~\ref{fig:critical}j}). Finally, higher-order time derivatives of the growth rate \( \hat{g}(\tau) \) peaked sharply near the transition point \( \tau^{\dagger} \) (\textbf{Fig.~\ref{fig:critical}d}), marking a sudden change in dynamics. These features align with classical signatures of a dynamical phase transition, suggestive of critical-like dynamics underlying the coordination of pattern growth and arrest.

We next asked whether these features were present in experimental data. Tracking the Bra$^+$ mesodermal domain in over 500 colonies (\( N = 520 \)) across a wide range of radii (\( R = 50\text{--}400\, \mu\mathrm{m} \)), we found growth curves with similar saturation behavior as in simulations (\textbf{Fig.~\ref{fig:critical}e}). Notably, the Bra$^+$ domain size decreased beyond \( t \approx 40 \, \mathrm{h} \), likely reflecting biological processes such as differentiation or signaling decay. As our focus is on domain expansion, we restrict analysis to the monotonic growth phase prior to this point.

As in simulations, experimental growth curves collapsed when normalized by their maximum pattern size \( S_p^{\max} \), showing remarkable alignment across colony sizes (\textbf{Fig.~\ref{fig:critical}f}). Moreover, the final pattern area scaled with colony radius as \( S_p^{\max} \sim R^{\epsilon} \), with exponent \( \epsilon = 1.92 \pm 0.19 \), matching the simulation-derived value and reinforcing the scale-invariant nature of mesodermal expansion.

Analyzing the relative growth rate \( g(t) \), we again identified an early power-law decay followed by an exponential tail (\textbf{Fig.~\ref{fig:critical}g}). The critical transition between these regimes occurred at \( t^{\dagger} = 22\text{--}27 \, \mathrm{h} \), consistent with simulation predictions when scaled using \( k_A = 0.14 \, \mathrm{h}^{-1} \) (\textbf{Fig.~\ref{fig:critical}h}). As in simulations, this transition was marked by an increase in the first and second derivative of the growth rate, supporting the existence of a dynamical phase transition.

We extracted the growth exponents \( \gamma \) and \( \beta \) from both simulations and experimental data (\textbf{Fig.~\ref{fig:critical}i,j}). In simulations, \( \gamma \) values increased with colony size and were slightly higher in the source region than the pattern, while \( \hat{\beta} \) remained invariant. Experimental values were consistent with simulations: \( \gamma \) increased with \( R \) (albeit with marginal significance, \( \rho = 0.69 \), \( p = 0.058 \)), and \( \beta \) was independent of size. Notably, the simulated exponential decay rates cluster around \( \hat{\beta} \approx 0.25 \), and when rescaled using the experimental degradation rate \( k_A = 0.14 \, \mathrm{h}^{-1} \), the resulting physical units \( \beta \) partially overlap with the distribution of experimental values for certain system sizes, although they tend to fall outside the central range of simulated values.

Altogether, our results suggest that in 2D gastruloids, critical-like dynamics may support size-aware growth control, enabling symmetry-breaking, scaling, and growth arrest to emerge in a coordinated, self-organized manner. 

\section*{Discussion}

Our study demonstrates that 2D gastruloids, organized PSC colonies cultured in adherent conditions, exhibit intrinsic self-organizing dynamics that model key features of early development, including spontaneous symmetry-breaking and mesodermal axis formation.  Notably, these phenomena arise in the absence of imposed gradients or extraembryonic tissues, challenging the view that such external cues are necessary initiators of polarity. Instead, our results support a model in which external signals may act secondarily to reinforce or stabilize emergent intrinsic order~\cite{morgani_micropattern_2018}. A key insight from our work is that symmetry-breaking is not simply a geometric event but is intrinsically coupled to system size. As colonies increase in radius, the probability and nature of symmetry transitions change, implicating a collective sensing mechanism that links local fate decisions to global geometry. 

We uncovered a universal, time-resolved scaling behavior: the mesodermal domain grows via a sublinear power-law before undergoing a size-dependent arrest governed by exponential decay. This biphasic growth curve was consistent across experimental and simulated systems of varying sizes, and revealed two dynamical transitions: one marking the onset of self-amplifying expansion (\( t^* \)), and another signaling growth arrest (\( t^{\dagger} \)). These transitions coincide with changes in higher-order time derivatives of the growth rate, suggesting the presence of a transient critical regime. Such critical-like dynamics—exhibiting slowing down and increased sensitivity near \( t^{\dagger} \)—are rare in biological patterning systems, but may provide a general design principle for coordinating fate specification with growth~\cite{mora_are_2011, munoz_colloquium_2018}. Recent studies in 3D gastruloids report spatial scaling of patterned regions~\cite{oriola_cell-cell_2024}, but lack a mechanistic account of growth dynamics. Future work could apply our framework to test whether similar critical transitions underlie self-arrest in 3D systems. Additionally, integrating recent findings on tissue flows that enhance asymmetry~\cite{gsell_marangoni-like_2025} may reveal how mechanical and biochemical cues co-regulate symmetry-breaking.

Our minimal model, a two-component Turing system with reactive boundaries, recapitulates the spatial and temporal features observed in the experiments and offers a novel mechanism for linking positional information to internal dynamics. In particular, the model suggests that fate domains may be governed not by morphogen concentration alone, but by the dynamics of morphogen \textit{production}. This distinction offers a resolution to longstanding debates between threshold-based versus source-based mechanisms of cell differentiation\cite{aguilar-hidalgo_critical_2018,ho_dynamics_2024}. Future experiments that directly manipulate or visualize morphogen production, such as inducible secretion systems and ligand biosensors, could test whether production dynamics serve as instructive cues for cell fate. By coupling growth arrest to source dynamics, the model introduces a feedback loop in which the patterning process itself modulates its own expansion, yielding robust, geometry-aware outcomes. A key test of the model will be to determine whether known signaling ligands in gastruloids, such as BMP4, Wnt, or Nodal, establish dynamic source regions consistent with the predicted spatial and temporal patterns. 

These findings carry implications for both developmental biology and synthetic morphogenesis. The ability of a system to self-arrest its growth in a size-dependent manner, without external timing cues or boundary imposition, suggests a powerful strategy for engineering tissues that scale appropriately with system size. Organoid platforms designed to exploit such feedbacks could achieve controlled morphologies, internally regulated growth phases, or spatial patterning that adapts to geometry. Moreover, the emergence of critical-like transitions in this context raises intriguing questions: Are such transitions a universal feature of self-organizing developmental systems? Can they be harnessed to control fate dynamics or amplify responsiveness near decision points?

By revealing how symmetry-breaking, scaling, and growth control emerge from a unified, self-organized mechanism, this work provides a conceptual framework for understanding how multicellular systems integrate geometry, signaling, and time. 


%

\end{document}